\documentclass{PoS}
\usepackage[utf8]{inputenc}

\usepackage[numbers]{natbib}
\usepackage{graphicx}
\usepackage{amsmath,amssymb,amstext}
\usepackage{subfigure}

\usepackage{float}

\usepackage{lmodern}

\usepackage{ etoolbox, xparse}
\usepackage{mathtools,nccmath}%

\usepackage{environ}

\newcommand{\rj}{R_{\rm jet}}

\newcommand{\geff}{\Gamma_{\rm eff}}

\usepackage{sidecap}
\sidecaptionvpos{figure}{c}

\title{\emph{Espresso} Acceleration of Ultra-High-Energy Cosmic Rays up to the Hillas Limit in Relativistic MHD Jets}

\ShortTitle{\emph{Espresso} in MHD jets}

\author{\speaker{Rostom Mbarek}\\
        University of Chicago, 5640 S Ellis Ave., Chicago, IL 60637 (USA)\\
        E-mail: \email{rmbarek@uchicago.edu}}
        
        \author{{Damiano Caprioli}\\
        University of Chicago, 5640 S Ellis Ave., Chicago, IL 60637 (USA)}

\abstract{\emph{Espresso} is a novel acceleration model for Ultra-High-Energy Cosmic Rays (UHECRs), where lower-energy CRs produced in supernova remnants experience a one-shot reacceleration in the relativistic jets of powerful Active Galactic Nuclei (AGNs) to reach energies up to $10^{20}$\,eV.
To test the \emph{espresso} framework, we follow UHECR acceleration bottom-up from injection to the highest energies by  propagating 100,000 particles in realistic 3D magneto-hydrodynamic (MHD) simulations of ultra-relativistic jets.
We find that simulations agree well with analytical expectations in terms of  trajectories of individual particles.
We also quantify that $\sim$~10\% of CR seeds gain a factor of $\sim\Gamma^2$ in energy, where $\Gamma$ is the jet's effective Lorentz factor;
moreover, about $0.1\%$ of the particles undergo two or more shots to achieve gains in excess of $\Gamma^2$.
Particles are generally accelerated up to the jet's Hillas limit, indicating that the \emph{espresso} mechanism should boost galactic CRs to UHECRs in typical AGN jets. 
Finally, we  discuss how \emph{espresso} acceleration in AGN jets is consistent with UHECR spectra and chemical composition, and also with the UHECR arrival directions measured by Auger and Telescope Array.}

\FullConference{36th International Cosmic Ray Conference -ICRC2019-\\
		July 24th - August 1st, 2019\\
		Madison, WI, U.S.A.}

\begin{document}

\maketitle

\section{Introduction}
The origin of the highest-energy cosmic rays (CRs) is one of the most prominent unresolved questions in astrophysics. 
Below $10^{17}$\,eV, diffusive shock acceleration (DSA) in supernova remnants (SNRs) is thought to be the main CR acceleration mechanism \citep[e.g.,][]{bell78a,bo78,bv07,pzs10, DSA}. As for Ultra-High-Energy Cosmic Rays (UHECRs) with energies between $\sim 10^{18}$~eV and $\sim 10^{20}$~eV, their sources and acceleration mechanism remain much less clear. Based on energetics and luminosity arguments, $\gamma$-ray bursts  \citep[e.g.,][]{vietri95,waxman95}, 
tidal disruption events  \citep[e.g.,][]{fp14},
newly-born millisecond pulsars \citep[e.g.,][]{blasi+07,fang+12}, and active galactic nuclei (AGNs) \citep[e.g.,][]{ostrowski00,murase+12} are potential UHECR sources. 
However, very often UHECR models are limited to back-of-the-envelope estimates of the maximum achievable energy for a given astrophysical object, rather than putting foward bottom-up mechanisms.

In reference \cite{espresso}, it has been suggested that UHECRs are produced in relativistic AGN jets through a general mechanism dubbed \textit{espresso} acceleration. 
Essentially, CR \emph{seeds} accelerated up to $10^{17}$\,eV in SNRs penetrate into a relativistic jet and receive a boost of a factor of $\sim \Gamma^2$ in energy, where $\Gamma$ is the Lorentz factor of the relativistic flow. One \emph{espresso} shot can boost the energy of galactic CRs by a factor $\Gamma^2\gtrsim 10^3$, transforming CRs at $10^{17}$~eV to UHECRs at $10^{20}$~eV, provided that $\Gamma\gtrsim 30$, as inferred from multi-wavelength observations of powerful blazars \citep[e.g.,][]{tavecchio+10,zhang+14}. 
Note that no assumptions are made on particle pitch-angle scattering (diffusion) or on the properties of the underlying magnetic turbulence, unlike in stochastic models that rely heavily on repeated acceleration at the jet interface and on magnetic turbulence at Larmor-radius scales \citep[e.g.,][]{ostrowski98,ostrowski00,kimura+18,fm18}.

The \emph{espresso} scenario has been corroborated with analytical calculations of CR trajectories in idealized jets, which confirm that most of the trajectories lead to $\sim \Gamma^2$ boosts regardless of the radial and longitudinal jet structures \citep{caprioli18,mc19}. 
The maximum energy achievable depends on the transverse ($\rj$) and longitudinal ($H$) sizes of the region with Lorentz factor $\Gamma$;
given a particle with initial Larmor radius $\mathcal{R}_{\rm i}$ and final radius $\mathcal R_{\rm f}\simeq \Gamma^2\mathcal R_{\rm i}$, such constraints read:
\begin{equation}\label{eq:hillas}
    \mathcal R_{\rm i} \lesssim\frac{\rj}{2\Gamma^2}; \quad
    \mathcal R_{\rm i} \lesssim \frac{2}{\pi}\frac{H}{\Gamma^3}
    \approx \frac{H}{\Gamma^3} 
    \quad \to \quad
    \mathcal R_{\rm f} \lesssim\frac{\rj}{2}; \quad
    \mathcal R_{\rm f} \lesssim  \frac{H}{\Gamma}
\end{equation}
Both conditions can be interpreted as the \emph{Hillas criterion} \citep{hillas84, cavallo78} in the transverse and longitudinal directions, where the latter is Lorentz-contracted.

However, analytical calculations cannot answer some fundamental questions such as: \emph{Is it possible to undergo more than one shot and thus exceed the $\Gamma^2$ gain? Is acceleration up to the Hillas limit generally achievable? What is the fraction of CR seeds that can undergo \emph{espresso} acceleration? Are reaccelerated particles beamed along the jet or are they released isotropically?}
To address these points, we traced test particles in a self-consistent 3D magneto-hydrodynamic (MHD) simulations of ultra-relativistic jets. 
We show some representative particle trajectories in \S\ref{sec3} and in \S\ref{sec5} we discuss the energy spectrum and angular distribution of the accelerated particles.
We conclude by summarizing the implications of our results for the origin of UHECRs in AGN jets (\S\ref{sec6}).

Throughout the paper, we denote quantities in the laboratory and flow frames respectively with $Q$ and $Q'$, and initial/final quantities with the subscripts~$_{\rm i} / _{\rm f}$; and introduce the particle gyroradius normalized to the jet radius, $\alpha$, such that $ \alpha ' \equiv \frac{\mathcal R '}{\rj} = \Gamma^2 \alpha_{\rm i}=\alpha_{\rm f}.$
We also consider two possible orientations of the jet toroidal magnetic field, $B_\phi\lessgtr 0$, which corresponds to the jet current along $z$ being $J_z\lessgtr 0$ and call them case A and B, respectively.

\section{Propagation in a full MHD simulation}\label{sec3}
In order to properly capture all the properties of a realistic astrophysical jet, we performed 3D simulations with  PLUTO \citep{mignone+12}, a massively-parallel relativistic MHD code that includes adaptive mesh refinement. 
Particles are then propagated within the jet using the relativistic Boris algorithm \citep[e.g.,][]{bl91}, which ensures long-term stability of the orbits. 

\subsection{Particle Trajectories}\label{par_prop}

\begin{figure*}
\centering
\includegraphics[width=0.48\textwidth, clip,trim= 0 0 0 0]{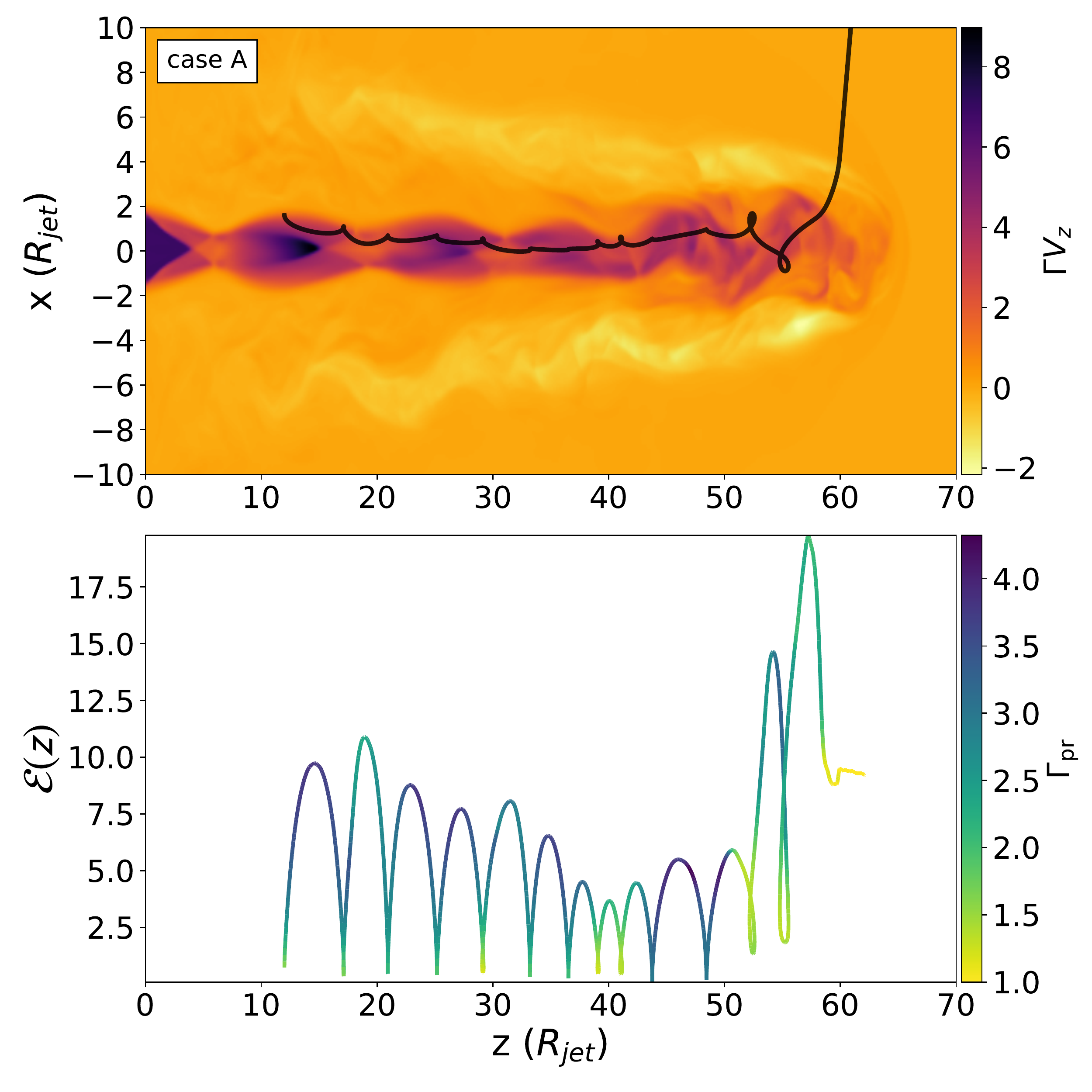}
\includegraphics[width=0.48\textwidth, clip,trim= 0 0 0 0]{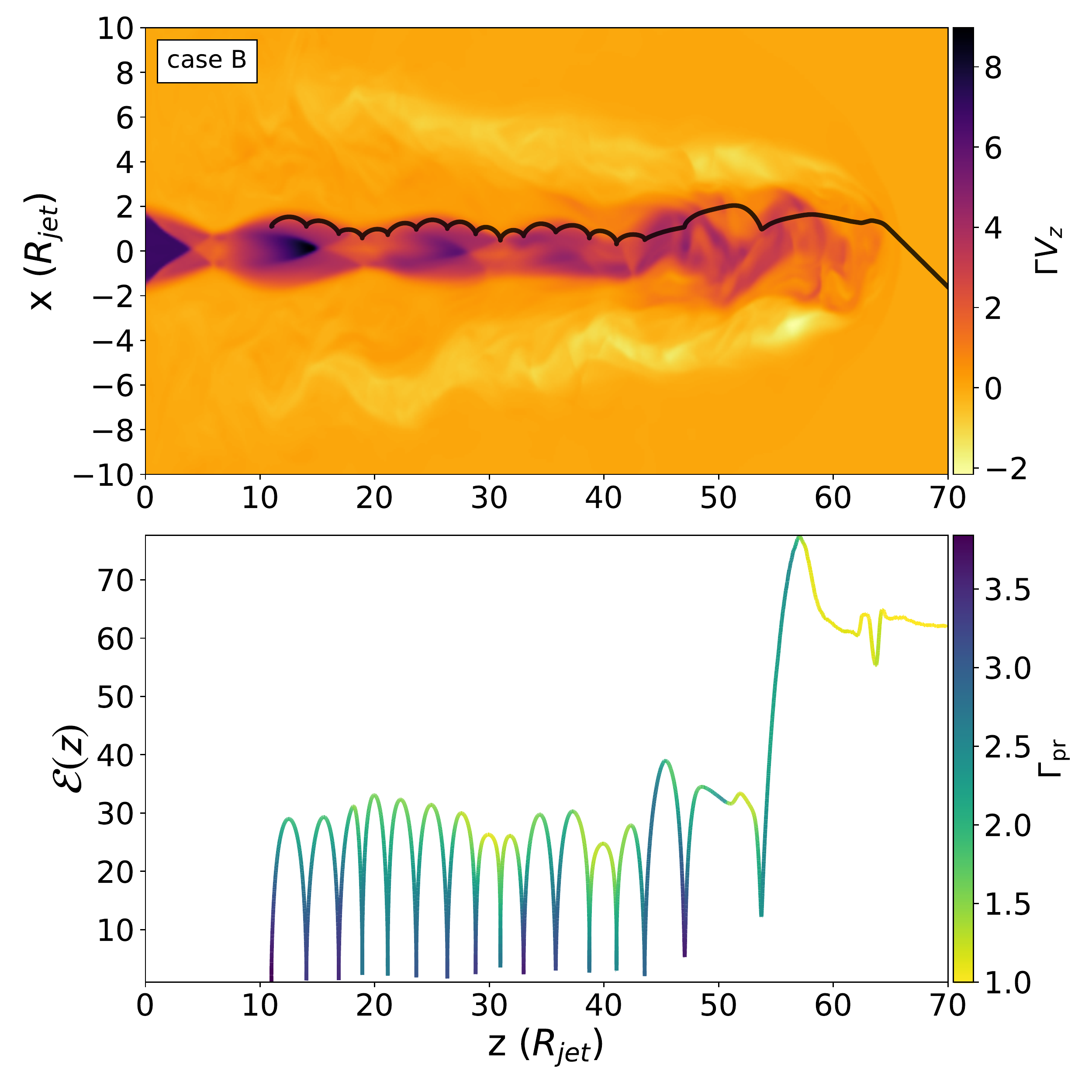}
\caption{Trajectory and energy gain for representative particles in case A and B (see \cite{mc19} for more details).
Top panels: particle trajectories overplotted on the 4-velocity component $\Gamma v_z$ of the flow.
Bottom panels: energy evolution as a function of $z$, color coded with the instantaneous Lorentz factor probed, $\Gamma_{\rm pr}$. 
Both particles are initialized with gyroradius smaller than the jet radius both gain energy in excess of $\Gamma_{\rm pr}^2$ up to the jet's Hillas limit via two \emph{espresso} shots.}
\label{traj}
\end{figure*}

For both case A and B, we propagate $\sim$100,000 protons with a broad range of initial gyroradii $\mathcal R$ and positions. Nuclei with charge $q=Ze$ have trajectories that can be derived by considering protons with the same rigidity $\rho\equiv E/q$ and with Larmor radii $\mathcal R= \frac{E}{ZeB}=\frac{\rho}{B}$. 
Protons are initialized with normalized Larmor radii spaced logarithmically in the interval $\alpha_{\rm i} = \frac{\mathcal{R_{\rm i} }}{\rj}\in [10^{-3.6}, 8]$ to cover a wide range of initial rigidities.
The magnitude of the magnetic field in the jet spine is larger than $B_0$:
averaging over the regions where $\Gamma\geq2$ returns $B_{\rm eff}\sim 7.2B_0$, so the effective Hillas condition (equation \ref{eq:hillas}) is satisfied for $\alpha_{\rm H}\sim 7.2$.

Particle trajectories in the MHD jet show many features common with those discussed in \citep{mc19} for simplified jets.
Figure~\ref{traj} illustrates two representative examples of \textit{espresso}-accelerated particles for case A and B.
The top panels show the particle trajectories overplotted on a 2D slice of the $z$ component of the flow 4-velocity, while the bottom panels show their energy gain $\mathcal{E}\equiv E_{\rm f}/E_{\rm i} $ as a function of $z$; the color code indicates the instantaneous Lorentz factor that they probe, $\Gamma_{\rm pr}$. These paradigmatic particle trajectories show that, regardless of the sign of the motional electric field:
i) particles gain energy because of ordered \emph{espresso} acceleration and not because of stochastic/diffusive processes in the cocoon; 
ii) it is possible to have $\mathcal{E}>\Gamma_{\rm pr}^2$, especially when kinks in the jet allow multiple acceleration cycles; 
iii) particles tend to gain energy up to the Hillas limit.

\subsection{Energy Distribution of the Accelerated Particles}\label{spectra_section}
\begin{figure*}
\centering
\includegraphics[angle=0,scale=.4,clip=true, trim= 10 0 0 0]{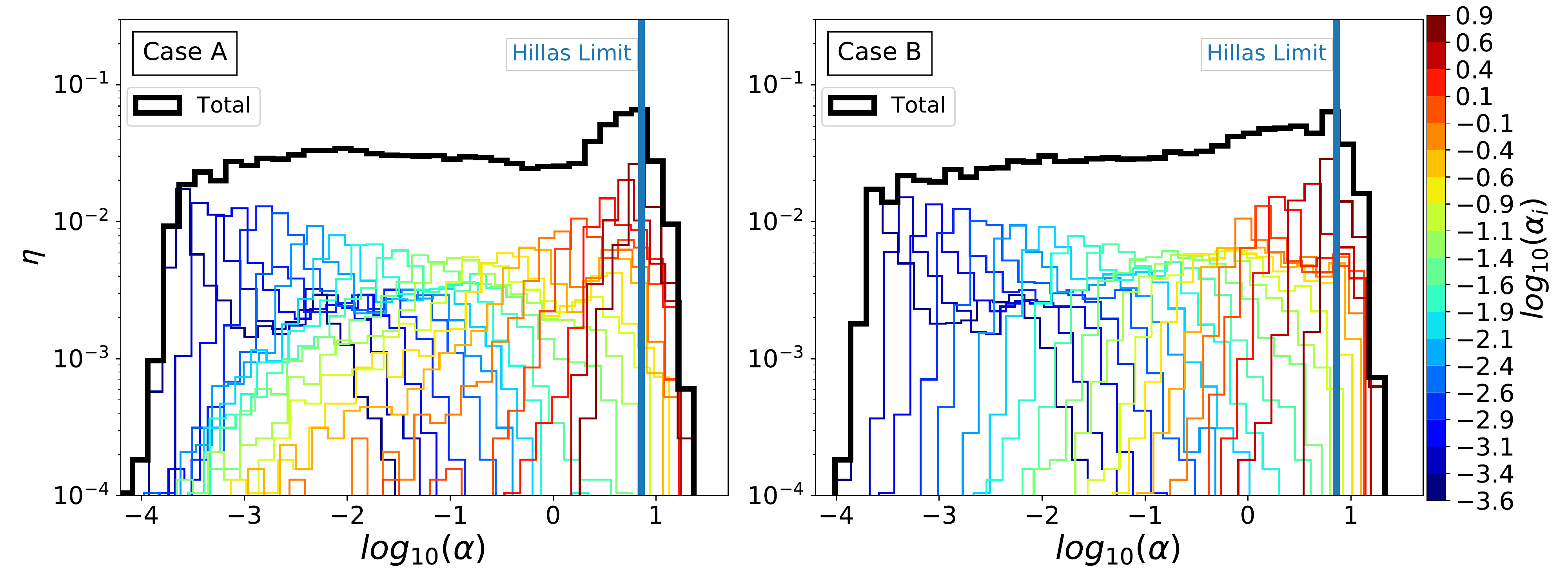}
\caption{Distribution of the Larmor radii of reaccelerated particles obtained for a flat injection spectrum, for both case A and B.
The black line shows the cumulative spectrum, while colored histograms correspond to different initial Larmor radii as in the color bar.
Seeds with $\alpha_{\rm i} \lesssim 1$ can undergo boosts as large as $\sim 50-100\gg\geff^2$, where $\Gamma_{\rm eff}$ is defined as the mean value of $\Gamma$ in the relativistic regions. 
For $\alpha_{\rm i} \gtrsim 1$ the energy gain is smaller and saturates at $\alpha_{\rm H}\approx 8$ (Hillas criterion).
}
\label{flat_spec}
\end{figure*}

Figure~\ref{flat_spec} shows the overall spectrum of emerging particles (solid black line), divided in the spectra produced by particles with different initial $\alpha_{\rm i}$ (colored histograms) in our simulated jet.
Four features are of interest: 
i) low-energy particles can go through multiple \emph{espresso} shots and gain energy beyond $\Gamma^2$; 
ii) the fraction of particles that are reaccelerated decreases at both ends of the  $\alpha_{\rm i}$ range; the optimal range for reacceleration is $\alpha_{\rm i}\in [\sim 0.01,\sim 1]$;
iii) the cumulative spectrum is truncated at $ \alpha_{\rm H} \sim 7.2$, the jet's Hillas limit;
iv) particles pile up close to the Hillas limit which results in a spectrum flatter than the injected one. 

\section{\emph{Espresso} Acceleration of UHECRs}\label{sec5}
Let us now consider the results above in the context of typical AGN jets and for different species in seed CRs. In the MHD simulation, Larmor radii are normalized to the jet radius and the initial magnetic field. If we set  $\rj \sim 15$ pc and $B_0\sim 1\mu$G, we associate physical rigidities to the injected particles;
the rigidity of the knee, $\rho_{\rm knee}\simeq 3\times 10^6$~GV, would correspond to $\alpha \simeq 0.2$.
In this section we focus on particles with initial rigidities $\rho_{\rm i} \in 3\times [10^3, 10^{6}]$~GV. 
The energy spectra of CR seeds are parametrized as in \cite{espresso}.

\subsection{Reacceleration Efficiency}
\begin{SCfigure}[][h]
  \centering
  \caption{\protect\rule{0ex}{-5ex} Cumulative distribution of energy gains of particles with $\rho_{\rm i}\in 3\times [10^3,10^6]$~GV.
Upper curves correspond to particles initialized in the whole domain, while lower curves consider only particles initialized in regions with $\tau\leq 0.1$, where $\tau$ is a tracer of the relative local abundance of jet/ambient material; $\tau=1$ and $\tau=0$ indicate pure jet and ambient material, respectively.}
  \includegraphics[ width=0.5\textwidth]%
    {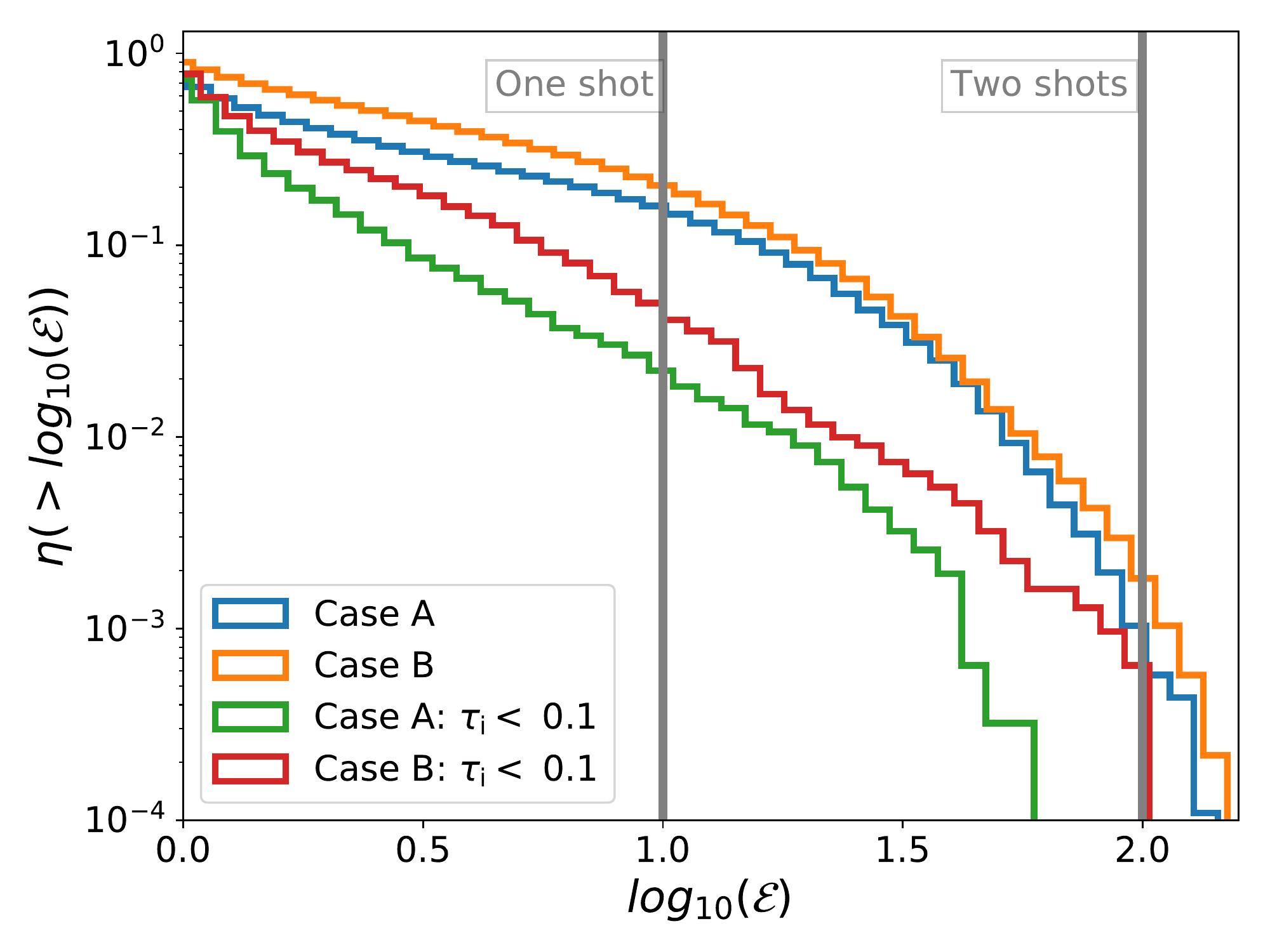}%
\label{rat_spec_cumul}

\end{SCfigure}

Figure~\ref{rat_spec_cumul} shows the distribution of the final energy gains of particles with $\rho_{\rm i}\in 3\times[10^3,10^6]$ GV, for both case A and B. 
We find that $\sim 38 \%$ ($\sim 53 \%$) of case~A(B) particles gain at least a factor of 2 in energy, and about $\sim 14 \%$ ($\sim 18 \%$) of case~A(B) particles gain a factor of $\geff^2$, where $\Gamma_{\rm eff}$ is defined as the mean value of $\Gamma$ in the relativistic regions. 
Also, $\sim 0.06 \%$ ($\sim 0.18 \%$) of the particles achieve an energy gain of 100 or more in case A(B), corresponding to two full \emph{espresso} shots or $\mathcal E\gtrsim \geff^4$. 
Particles that do not gain much energy either pass through the spine of the jet without gyrating because of their large rigidity, or cannot enter the flow because their initial Lorentz factor was too small.

\subsection{Spectrum and Elemental Composition of the Reaccelerated Particles}
\begin{figure*}
\centering
\includegraphics[width=\textwidth,clip, trim= 0 0 0 0]{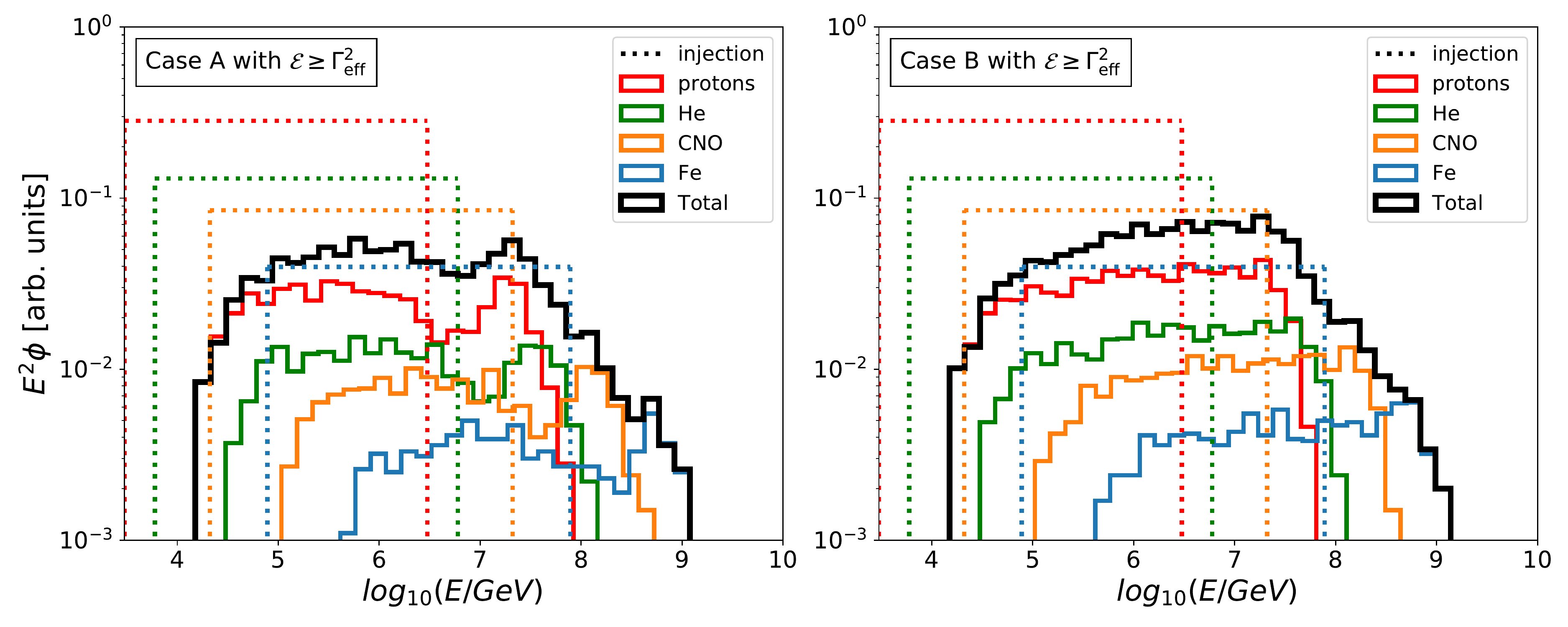}
\caption{Energy spectrum of the particles that experience at least one \emph{espresso} shot ($\geff^2$ gain) and escape the jet, assuming the injection spectrum from \cite{espresso} (dotted lines). }
\label{rig_spec}
\end{figure*}

The \emph{espresso} model predicts that the chemical composition observed in galactic CRs, which is increasingly heavy above $10^{13}$~eV \citep[e.g.,][]{hoerandel+06,ku12}, should be mapped into UHECRs. 
Figure~\ref{rig_spec} shows the energy spectrum produced in our calculations, considering only particles that gained at least $\mathcal E\gtrsim \geff^2$.
The cutoffs of the CR seeds are boosted up by a factor of $\geff^2\sim 10$ with a high-energy tail due to particles that experienced multiple shots. 
Below these cutoffs, spectra of outgoing particles are slightly harder than the seed ones because particles pile up at the Hillas limit;
the spectrum of accelerated particles flattens also at low energies because particles with $\alpha_{\rm i}\ll 1 $ are barely boosted.

\subsection{Angular Distribution of the Released Particles}\label{sec:anisotropy}

A potential correlation between the UHECR directions of arrival and the local AGNs is intimately connected to the release of particles from their sources, that is, whether reaccelerated particles are beamed along the jet or not. 
In the former case, UHECRs would trace AGNs with jets that point at us (blazars and flat-spectra radio quasars), while in the latter case all radio-loud AGNs may contribute, generating a more isotropic signal. 
In our simulations we observe that the geometry of the jet's toroidal magnetic field regulates how accelerated particles are released: case-A particles escape quasi isotropically, while case-B ones are beamed along the jet axis.
This dichotomy is caused by the radial electric field in the cocoon, $E_r=v_z B_\phi$, which reverses its sign in the two cases; 
the direction of $B_\phi$ in the jet goes in hand with a radial electric field that either scatters ($E_r>0$, case A) or collimates ($E_r<0$, case B) the particles that escape the jet.

\section{Conclusions}\label{sec6}

We have extensively analyzed the acceleration of UHECRs in relativistic AGN jets by propagating test-particles in both synthetic jet structures and full 3D relativistic MHD simulations. 
In our simulations, particles are accelerated up to the maximum theoretical energy (the Hillas limit) via ordered \emph{espresso} cycles, with no contribution from stochastic acceleration due to the velocity shear at the jet/cocoon interface, turbulent acceleration in the cocoon, or repeated diffusive shock acceleration in the AGN lobes \cite{kimura+18,osullivan+09,matthews+19}.
Within the \emph{espresso} framework, which relies on very general assumptions and is verified in bottom-up simulations, all of the UHECR observables (spectrum, maximum energy, energetics, composition, anisotropy) can be accounted for. 
The reader can refer to \cite{mc19} for a more extended discussion of the type of AGNs that may contribute the most to the UHECR flux.

Some quantitative conclusions from our simulation campaign are:
\begin{itemize}
\item 
$\sim$~10\% of the injected particles are boosted by a factor of $\mathcal E\gtrsim \geff^2$ in energy (one \emph{espresso} shot) and $\sim 0.1\%$ of them gain $\sim \geff^4$  (two \emph{espresso} shots).
These values are larger than the $\gtrsim 10^{-4}$ efficiency necessary to sustain the UHECR flux \cite{espresso}.

\item Since multiple shots are possible (and favored by jet wobbling and kinks) and since the energy gain scales as $\mathcal{E}\sim \geff^{2N}$, where $N$ is the number of shots, a few \emph{espresso} cycles are generally sufficient to reach the Hillas limit even in AGNs with moderate Lorentz factors (e.g., $N\lesssim 3$ for $\Gamma\gtrsim 3$).

\item The angular distribution of the reaccelerated particles depends on the direction of the radial electric field. 
In case A ($B_\phi<0$) particles are released quasi isotropically, while in case B ($B_\phi>0$) particles are preferentially beamed along the jet axis.
Since the sign of $B_\phi$ likely depends on the details of how matter is accreted on the black hole, UHECRs are not necessarily beamed as the jet's $\gamma$-ray emission.

\item These results, plus some energetics considerations \cite{mc19}, suggest that radio-loud AGNs with extended jets (Fanaroff-Riley II) are expected to be prominent UHECR sources. 

\end{itemize}

\bibliography{Total_JR}
\end{document}